\documentclass{aa}  
\DeclareUnicodeCharacter{2061}{}
\usepackage[varg]{txfonts}
\usepackage[export]{adjustbox}
\usepackage{tabularx}
\usepackage{multirow}
\usepackage{array}
\usepackage{amsmath}
\usepackage{chemformula}
\usepackage{textcomp}
\usepackage{amsmath}
\usepackage{placeins}
\usepackage{float}
\usepackage{natbib}
\usepackage{lscape}
\usepackage{xcolor}
\usepackage{stfloats} 
\usepackage{float}
\usepackage{soul}
\usepackage{xcolor}
\usepackage{graphicx}
\usepackage{txfonts}
\usepackage{lipsum}
\usepackage{subcaption}         % necessary for continued figures, example in section 3
\usepackage{lscape}             % to rotate a single page table, example in appendix.
\usepackage{placeins}           % useful with \FloatBarrier, to keep 
\usepackage[breaklinks=true]{hyperref} %% to avoid \citep line fills
\bibpunct{(}{)}{;}{a}{}{,}             %% natbib format for A&A and ApJ
\usepackage{xcolor}
\begin{document}

%%%%%%%%%%%%%%%%%%%%%%%%%%%%%%%%%%%%%%%%
% if you use custom commands in your title,
% ensure to check your title when submitting!
%%%%%%%%%%%%%%%%%%%%%%%%%%%%%%%%%%%%%%%%
   \title{How nanoscale physics shapes ice formation in the Universe}

   \subtitle{Rethinking gas freeze-out on dust grains}

%%%%%%%%%%%%%%%%%%%%%%%%%%%%%%%%%%%%%%%%
% Please separate each author with the \and command
%
% Use the \corrauth to provide the corresponding
% author address. It will be automatically inserted as 
% footnote in the PDF output.
%
% Please DO NOT include ORCIDs next to author names.
% Instead, please provide an active address for each coauthor:
% it will be automatically extracted by EDPS editorial system, 
% and co-authors will be be able to authenticate their ORCID.
%
% Only authenticated ORCIDs will be taken into account.
% ORCIDs included here will be removed.
%%%%%%%%%%%%%%%%%%%%%%%%%%%%%%%%%%%%%%%%

   \author{Ph. Parent\inst{1}\corrauth{philippe.parent@univ-amu.fr}        % use \corrauth for the corresponding author
        \and C. Laffon\inst{1}\email{carine.laffon@univ-amu.fr} 
        \and S. Curiotto\inst{1}\email{stephano.curiotto@cnrs.fr}
        \and D. Ferry\inst{1}\email{daniel.ferry@cnrs.fr}
        \and C. Stadler\inst{1,2}\email{cstadler@phys.au.dk}
        \and F. G. Doktor\inst{2}\email{frederik_doktor@phys.au.dk}
        }

   \institute{ Aix Marseille Univ, CNRS, CINaM, Marseille, France
   \and Center for Interstellar Catalysis (InterCat), IFA, Aarhus University, Denmark}

   \date{\today}

% \abstract{}{}{}{}{}
% 5 {} token are mandatory
 
\abstract
% context heading (optional)
{In cold molecular clouds, gas freeze-out onto dust grains initiates the formation of interstellar ice mantles. CO and other heavy species are generally expected to accrete efficiently from the gas phase, and astrochemical models have commonly assumed near-unity sticking probabilities at low temperature. However, recent laboratory measurements on realistic grain analogues show that sticking can remain far below unity even at 10 K. These observations call for rethinking gas-grain interactions beyond the conventional hit-and-stick picture.}
% aims heading (mandatory)
{Using CO as a prototype for the low-temperature accretion of heavy species, we aim to identify the physical origin of reduced sticking and to determine how nanoscale surface morphology controls gas freeze-out and the onset of ice growth.}
% methods heading (mandatory)
{CO adsorption at 10 K was investigated on highly oriented pyrolytic graphite, used as a flat reference surface, and on carbon soot, used as a rough cosmic dust grain analogue. X-ray photoelectron spectroscopy (XPS), low-temperature scanning tunneling microscopy (STM), kinetic Monte Carlo simulations (KMC), and a thermodynamic description were combined to follow ice growth and relate molecular retention to local surface morphology.}
% results heading (mandatory)
{XPS measurements show that CO does not adsorb with unit sticking on graphite. Adsorption proceeds sequentially through initial monolayer growth, a reduced-retention crossover near monolayer completion, and delayed multilayer growth. STM reveals that CO is highly mobile at 10 K on graphite terraces and is effectively retained only after reaching stabilizing configurations, including island edges and terrace steps. On soot, the same sequence occurs at much higher exposures due to lower sticking coefficients, consistent with KMC simulations showing preferential retention in concave regions and poor wetting of convex asperities.}
% conclusions heading (optional)
{Low-temperature sticking is controlled by post-impact dynamics, where mobile species transiently explore the surface and desorb unless they reach a stabilizing site. Nanoscale morphology amplifies this physical mechanism, reducing effective sticking probabilities, delaying gas freeze-out, and strongly limiting the accretion of CO and, more generally, heavy species on realistic dust grains.}

\keywords{Astrochemistry -- Methods: laboratory: solid state -- Solid state: volatile}
   \maketitle
   \nolinenumbers % arXiv does not allow line numbers in the submitted PDF

%%%%%%%%%%%%%%%%%%%%%%%%%%%%%%%%%%%%%%%%%%%%%%%%%%%%%%%%%%%%%%
\section{Introduction}
Astrochemical models have long assumed that gas-phase species stick to interstellar dust grains with near-unity probability \citep{Cuppen2017, DHendecourt1985a}. This initial adsorption step governs grain-surface chemistry, from molecule formation to ice accretion, and is therefore central to astrochemistry. Yet the microscopic mechanism by which volatile species become stabilized on grains remains poorly understood.
Until recently, most experimental insights into this process came from adsorption studies on idealized flat substrates \citep{He2016}. Real interstellar grains, however, are rough submicrometer aggregates with complex morphologies known to influence gas accretion and surface chemistry \citep{Christianson2021, Cuppen2007, Herbst2006, Jones2016, Jones2017, Jones2019a, Potapov2025, Potapov2021}. Recent X-ray photoelectron spectroscopy measurements on realistic grain analogues, including carbonaceous and silicate materials, have shown that \ch{H2O}, \ch{CO2}, \ch{N2} and \ch{CO} exhibit sticking probabilities far below unity \citep{Laffon2021, Stadler2024}. Incorporating these reduced sticking coefficients into astrochemical models significantly alters gas-ice partitioning and improves agreement with observed abundances in molecular clouds \citep{Stadler2025}.
Why does sticking remain inefficient despite barrierless adsorption at the low temperatures characteristic of cosmic dust? Here we show that the answer lies in post-impact dynamics. After capture, adsorbates enter a transiently mobile state in which surface exploration competes with escape back to the gas phase. Even on atomically flat substrates, retention requires access to sparse stabilizing configurations. On realistic dust grains, nanoscale morphology amplifies this selectivity by confining retention to a limited subset of favorable environments, leaving much of the surface effectively bare even at 10 K. These observations lead to a thermodynamic framework in which sticking is not treated as the irreversible outcome of capture, but as a retention probability set during this exploratory phase. By linking local free-energy biases to molecular retention, this framework provides a physical basis for connecting nanoscale surface structure to surface reactions and ice accretion on cosmic dust across molecular clouds, protostellar envelopes, and protoplanetary disks.

\section{Methods}

X-ray photoelectron spectroscopy (XPS) experiments were conducted under UHV at 10 K using the SUMO setup at CINaM. The method used to extract the sticking coefficient \(S\) is described in detail in the Supplementary Information of \cite{Laffon2021}. Briefly, CO is dosed isotropically in a dedicated sub-chamber, and the incident molecular flux is derived from the background pressure using the Knudsen-Hertz expression. The substrates include a gold foil for reference (99.95\% Alfa Aesar), a freshly cleaved highly oriented pyrolytic graphite (HOPG), a soot sample composed of carbon nanoparticles (see below), and a non-porous, crystalline \ch{H2O} ice layer deposited on HOPG at 140 K, selected to suppress CO bulk diffusion and isolate surface processes \citep{He2019}. X-ray source-sample distance and source intensity are carefully adjusted to avoid CO photodesorption and photolysis. 
XPS spectra are acquired at a take-off angle of 55°, at which the average photoelectron path length through an overlayer on spherical particles equals that on a flat surface. This geometry enables direct comparison of overlayer intensities on rough and planar substrates \citep{Gunter1997,Gunter1995,Kappen2000}, a condition we have explicitly verified for soot and HOPG \citep{Laffon2021}. A normalization procedure accounts for surface-area differences and geometric shadowing of emitted photoelectrons, enabling extraction of sticking coefficients via a graphical method calibrated against the gold reference surface, for which CO forms a complete monolayer with unit sticking under identical conditions \citep{Laffon2021,Stadler2024}. Following the formalism described in \cite{Laffon2021}, the contribution of the adsorbate to the total XPS signal, \(I\)(CO), is obtained from the ratio of the peak areas:

\begin{equation}
\label{eqn1}
\begin{aligned}
I(\mathrm{CO}) &=
\frac{A(\mathrm{adsorbate})}
     {A(\mathrm{adsorbate}) + A(\mathrm{substrate})} \\
&=
\frac{A(\mathrm{C}\,1s) + A(\mathrm{O}\,1s)}
     {A(\mathrm{C}\,1s) + A(\mathrm{O}\,1s) + A(\mathrm{substrate})}.
\end{aligned}
\end{equation}

In Eq.~\ref{eqn1}, all peak areas correspond to fitted integrated areas corrected for their respective relative sensitivity factors. For the gold reference the substrate signal corresponds to Au~4f peaks, whereas for HOPG and soot it corresponds to the graphite C1s peak. Peak areas are obtained after background subtraction and fitting using CasaXPS. Sticking coefficients are obtained by comparing the exposures required to reach the same adsorbate intensity \(I\)(CO) on the investigated substrate and on the gold surface. Since CO adsorption on Au proceeds with unit sticking, the sticking coefficient on a given substrate is simply \(S = E_{\mathrm{Au}}/E\), where \(E_{\mathrm{Au}}\) and \(E\) are the exposures required to reach the same \(I(\mathrm{CO})\) intensity on gold and on the investigated substrate, respectively.

Carbon soot was generated from a pure paraffin candle \citep{Mulay2019}. It consists of fractal aggregates of turbostratic graphitic nanoparticles with a mean diameter of 25 ± 15 nm \citep{Laffon2021}. Particle morphology was characterized by transmission electron microscopy (TEM; JEOL JEM-2010, 200 kV, bright-field mode). Surface roughness was quantified from TEM images using a cross-sectional line-profile method to extract effective pore depth and width from 2D projections (Fig.~\ref{fig:soot_tem}). Upper and lower envelope lines were drawn through hilltops and valley bottoms to estimate characteristic pore depths (d =0.5 nm). Pore widths were determined from the full width at half maximum of individual valleys (w = 1.8 nm), with the surface area approximately evenly partitioned between concave pockets and convex asperities. Measurements were performed using Gwyddion 2.70.

Atomic-scale measurements were conducted at Aarhus University using low-temperature scanning tunneling microscopy (LT-STM) \citep{Doktor2025,Hornekaer2003}. CO was dosed from a gas line onto freshly cleaved HOPG, previously annealed at 1200 K, while the sample was kept at \(\sim11~\mathrm{K}\) during deposition. After dosing, the sample was cooled to 4 K for constant-current STM imaging. The quoted STM exposures are nominal values estimated from the background pressure measured in the STM chamber and are used only to define the relative dosing sequence. Because of differences in pressure gauges, dosing geometries, and temperature measurements, these STM exposure values and sample temperatures should not be regarded as directly equivalent to those used in the XPS experiments. STM image analysis was performed using Gwyddion 2.70.

Kinetic Monte Carlo (KMC) simulations were performed to model CO adsorption and diffusion on a flat surface and on a rough carbon substrate. Mobile CO adsorbates were represented as particles interacting on a static carbon lattice with hexagonal symmetry and periodic boundary conditions. Binding interactions were treated using nearest-neighbor counting, in which the local adsorption energy is determined solely by the number of occupied adjacent sites, a standard approach for simulating sub-monolayer adsorption and diffusion processes \citep{Combe2000, Ghosh2014, Krug1995, Pierre-Louis2007}. Gas exposure was simulated by maintaining a constant reservoir of CO above the substrate. Jump probabilities between sites followed Arrhenius-type rates with an attempt frequency of \(10^{13}\,\mathrm{s}^{-1}\). Rough morphologies were parameterized using pore depths and widths extracted from TEM images. Simulations were performed with T=10~K and CO--CO and CO--carbon interaction energies primarily set to \(E_{\mathrm{CO-CO}} = 900~\mathrm{K}\) and \(E_{\mathrm{CO-C}} = 1600~\mathrm{K}\), expressed in kelvin units, that is, as \(E/k_{\mathrm{B}}\). These values are consistent with the desorption temperature of pure CO ice (\(27~\mathrm{K}\), Fig.~\ref{fig:temperature_dependence}B) and with values commonly adopted in astrochemical compilations \citep{Minissale2022}. Additional simulations were performed using a reduced CO--carbon interaction energy, \(E_{\mathrm{CO-C}} = 900~\mathrm{K}\), to estimate the sensitivity of the kinetic behavior to substrate binding strength (Appendix A).

\section{Results}

To assess the impact of surface morphology on adsorption, we compare CO adsorption on graphite and carbon soot. HOPG provides a flat reference surface, whereas carbon soot provides highly corrugated surfaces with nanometer-scale concave pockets and convex asperities (Fig.~\ref{fig:soot_tem}), representative of the nanoscale roughness expected for interstellar grains. CO serves as a prototypical interstellar volatile for probing the fundamental mechanisms governing sticking on grains.
\subsection{X-ray photoelectron spectroscopy}
\begin{figure*}[htbp]
    \centering
    \includegraphics[width=14cm, height=10cm]{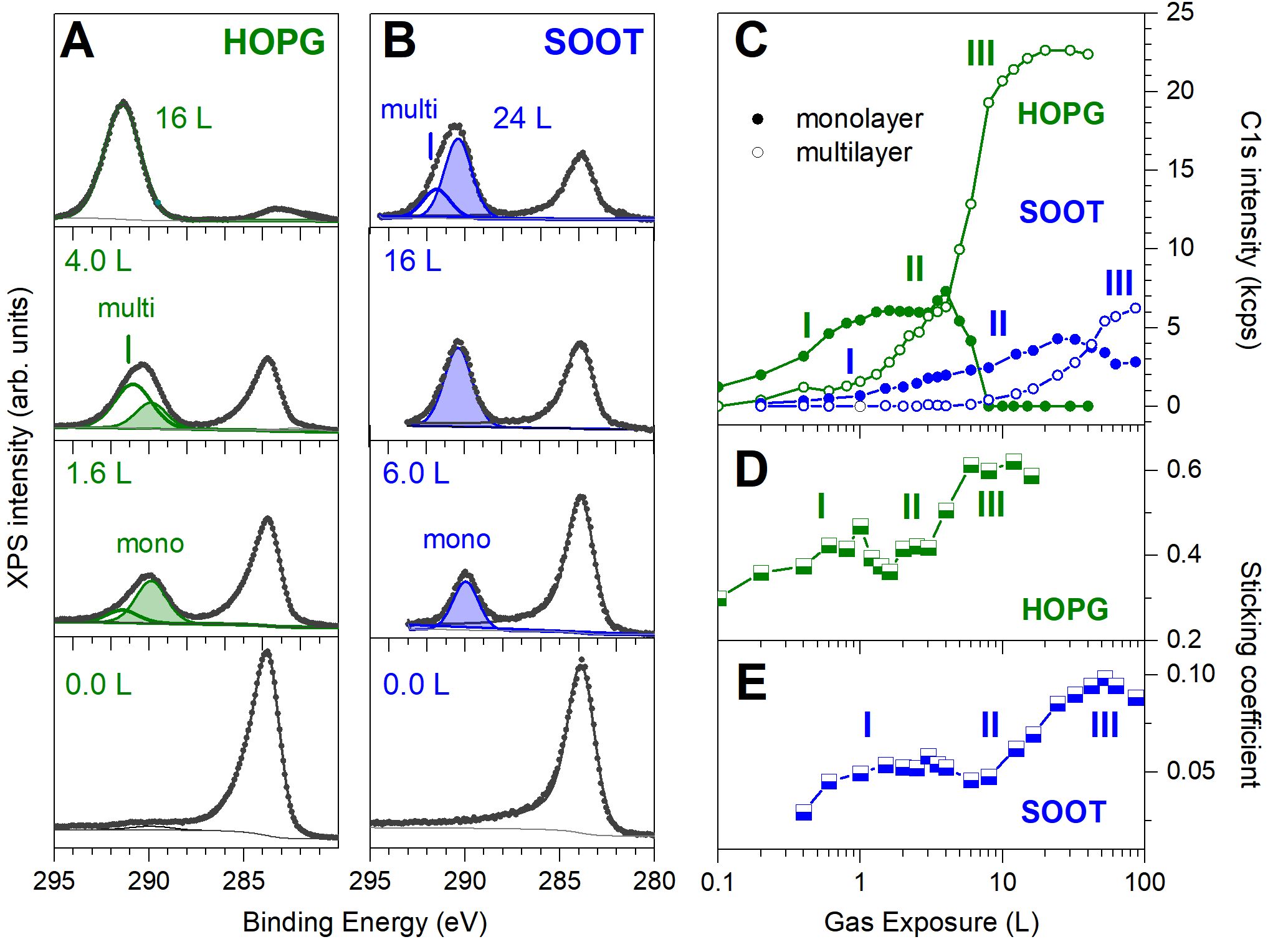}
\caption{XPS C1s signatures of CO adsorption on graphite and soot.
(A, B) Monolayer (\(\approx 290~\mathrm{eV}\)) and multilayer (\(\approx 291~\mathrm{eV}\)) C1s components for CO deposited on HOPG (green) and soot (blue). On HOPG, both components appear within a few Langmuir, whereas on soot the monolayer persists up to \(\sim16~\mathrm{L}\) and multilayer growth begins only beyond \(\sim20~\mathrm{L}\), indicating substantially reduced sticking.
(C) Evolution of monolayer and multilayer intensities with CO exposure, defining three regimes: Regime~I (monolayer growth), Regime~II (crossover near monolayer completion), and Regime~III (multilayer growth).
(D, E) Corresponding sticking coefficients \(S\) for HOPG and soot. Errors in \(S\) correspond to \(\pm 5\%\) for flat substrates and \(\pm 10\%\) for particles \citep{Laffon2021}.}
    \label{fig1}
\end{figure*}

XPS establishes the global adsorption behavior on both materials (Figure~\ref{fig1}A,B). With increasing CO exposure, a monolayer component appears near \(290~\mathrm{eV}\) in the C1s region, accompanied by attenuation of the substrate signal at \(284~\mathrm{eV}\). A second component appears at \(\sim291~\mathrm{eV}\) once the monolayer saturates, consistent with the onset of multilayer growth. On HOPG, both features appear within a few Langmuir. On soot, by contrast, the monolayer persists up to \(\sim16~\mathrm{L}\) and multilayer growth begins beyond \(\sim20~\mathrm{L}\), consistent with a substantially lower sticking coefficient \citep{Stadler2024}.

Exposure-dependent intensity traces (Fig.~\ref{fig1}C) reveal three regimes on both materials: initial growth (Regime~I), a crossover near monolayer completion (Regime~II), and multilayer growth (Regime~III). The XPS-derived sticking coefficient \(S\) (Fig.~\ref{fig1}D,E) follows the same sequence. In Regime~I on HOPG, \(S\) increases from \(\sim0.3\) to \(\sim0.5\) as monolayer growth approaches saturation, then decreases in Regime~II (\(\sim0.35\)), before rising again to \(\sim0.6\) in Regime~III. On soot, the same three regimes occur at higher exposures but with substantially lower sticking coefficients (\(0.03 \lesssim S \lesssim 0.1\)), indicating that effective stabilization is confined to a limited fraction of surface sites.

\subsection{Scanning tunneling microscopy}

\begin{figure*}[htbp]
    \centering
    \includegraphics[width=18cm, height=9cm]{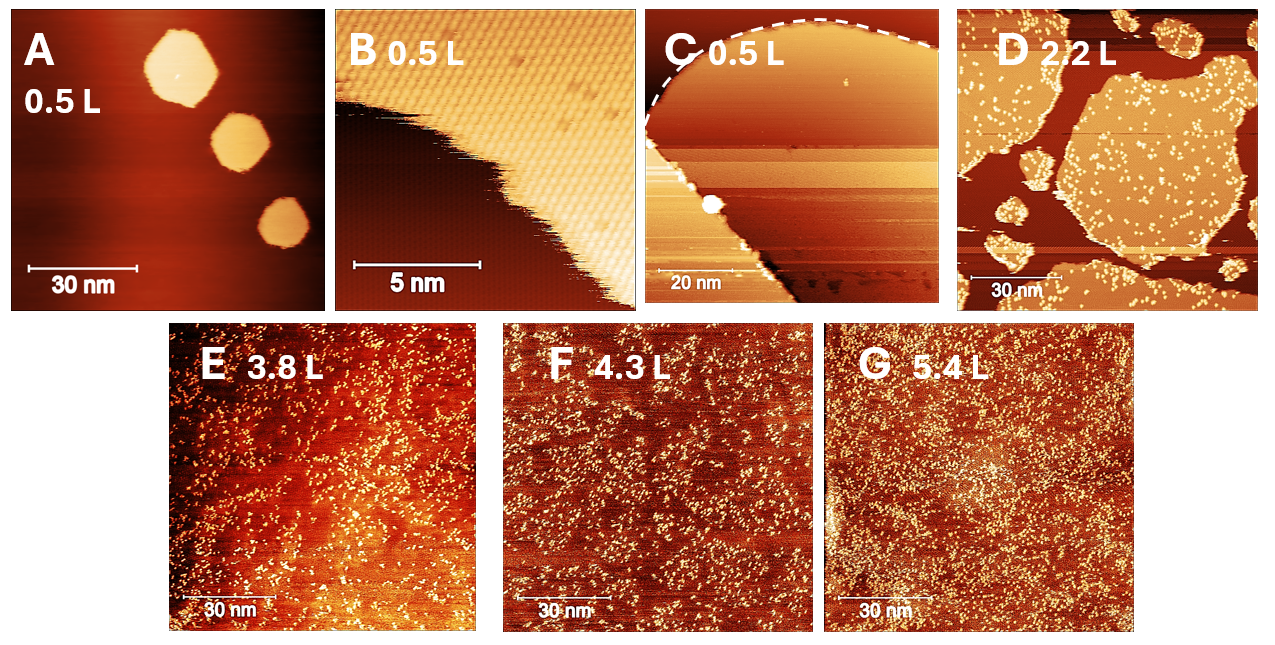}
\caption{STM images of CO adsorption on HOPG. (A--D) Regime~I. In (A,B), CO forms commensurate\(\sqrt{3}\) islands, while HOPG terraces remain free of stable monomers, indicating high mobility. Molecules diffuse across terraces until reaching stabilizing sites. In (C), a large island, indicated by the dashed line, is pinned at a terrace edge. In (D), islands expand and isolated second-layer nuclei appear as bright protrusions, corresponding to CO stabilized at favorable on-top sites. (E--G) Regime~II. At higher exposures, the density of these nuclei increases across the completed monolayer. Despite increasing exposure, extended second-layer domains do not form, indicating limited stabilization and hindered layer-by-layer growth. CO was dosed at \(\sim11~\mathrm{K}\), and images were recorded at \(4~\mathrm{K}\); scanning parameters: \(I = 63~\mathrm{pA}\), \(V = -2.1~\mathrm{V}\). }
    \label{fig2}
\end{figure*}

STM provides atomic-scale observation of first- and second-layer growth (Figure~\ref{fig2}). At low exposure (Regime I), CO forms commensurate \((\sqrt{3}\times\sqrt{3})R30^\circ\) hexagonal islands on HOPG terraces \citep{Belak1985, Harris1979, You1985} (Fig.~\ref{fig2}A,B; Fig.~\ref{fig:stm_linescans}), hereafter denoted \(\sqrt{3}\). Crucially, the absence of isolated monomers on HOPG terraces indicates that CO remains highly mobile after impact and explores the surface until it reaches a stabilizing configuration. This exploration promotes island growth and preferential accumulation at favorable sites. Large islands form at step edges, which act as efficient stabilization centers (Fig.~\ref{fig2}C). Preferential accumulation also occurs along smooth topographic modulations on HOPG (Fig.~\ref{fig:stm_soft_steps}), showing that such concave regions locally stabilize adsorbates. In Regime~I, stabilization is thus limited to sparse sites and most adsorbates escape after surface exploration, yielding an initial sticking coefficient of \(\sim0.3\). Rapid reconstruction after tip-induced perturbations at \(4~\mathrm{K}\) (Fig.~\ref{fig:stm_tip_reconstruction}) indicates that lateral CO--CO interactions dominate over weak CO--graphite binding, and consistent with high terrace mobility of CO monomers.

As islands expand with exposure, their edges provide additional stabilizing sites and sticking increases to \(\sim0.5\). Bright protrusions appear (Fig.~2D), corresponding to stabilized CO at favorable on-top sites on first-layer islands. At monolayer completion, the system enters the crossover Regime~II (Fig.~2E--G), where \(S\) decreases to \(\sim0.35\). Only a fraction of incoming CO is retained, and although the protrusions increase in density with increasing exposure, extended domains do not form.

\subsection{Kinetic Monte Carlo simulations}

\begin{figure*}[htbp]
    \centering
    \includegraphics[width=17cm, height=9cm]{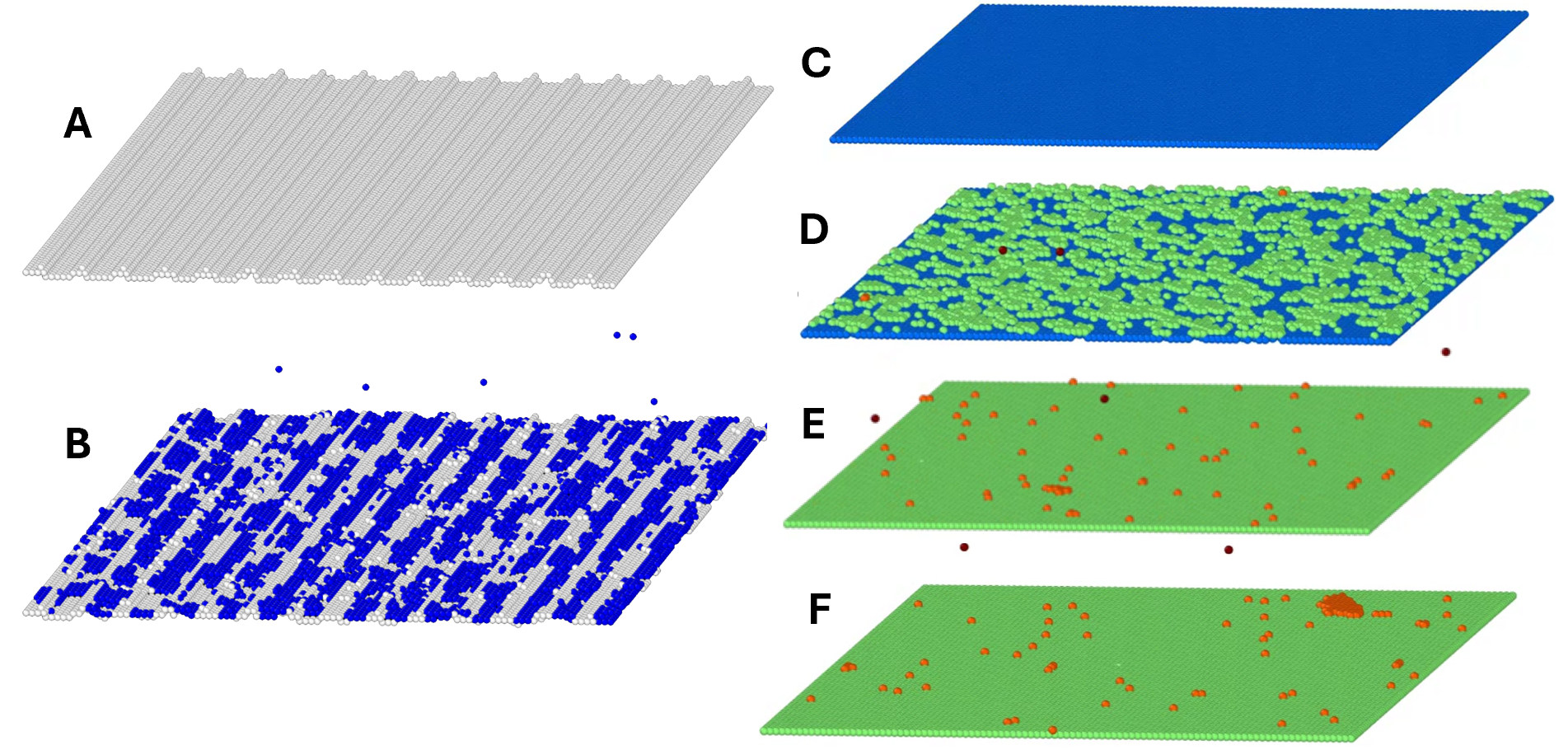}
\caption{Kinetic Monte Carlo simulations of CO adsorption on rough and flat carbon surfaces. (A, B) On a rough carbon substrate (grey), parameterized from TEM-derived soot morphology, CO (blue) preferentially accumulates within concave pockets, while convex regions remain effectively non-wettable. As a result, only a limited fraction of the surface contributes to retention, yielding a low global sticking coefficient. (C--F) On a flat carbon surface (blue), CO (green) forms first-layer islands that grow and merge into a defect-free layer. (E) On this ordered layer, additional CO (orange) appears only transiently: they represent instantaneous adsorbate positions and do not correspond to stabilized second-layer configurations. (F) Second-layer nucleation remains hindered until rare favorable configurations emerge, here in the top-right aggregate, after which a second-layer island develops. Simulations were performed with \(T = 10~\mathrm{K}\), \(E_{\mathrm{CO-CO}} = 900~\mathrm{K}\), and \(E_{\mathrm{CO-C}} = 1600~\mathrm{K}\). Movie 1 shows the corresponding flat-surface simulation, while Movie 2 shows the reduced \(E_{\mathrm{CO-C}} = 900~\mathrm{K}\) case discussed in Appendix A. }
    \label{fig3}
\end{figure*}

Numerical simulations illustrate the role of nanoscale topography. KMC simulations probe how surface morphology shapes nucleation and growth  (Figure~\ref{fig3}). On a rough carbon surface (Fig.~\ref{fig3}A,B), parameterized from TEM images of soot (Fig.~\ref{fig:soot_tem}), adsorbates preferentially accumulate within concave pockets, whereas convex asperities remain unwettable. Only a limited fraction of the surface supports ice growth, showing how curvature confines retention to favorable regions and promotes local multilayer buildup rather than lateral wetting of adjacent areas \citep{Christianson2021}.

On a flat graphite surface (Fig.~\ref{fig3}C--F), the simulated first-layer islands are irregular, unlike the hexagonal islands observed by STM. Their shape indicates diffusion-limited aggregation \citep{Witten1981}, caused by the limited adsorbate mobility in the KMC model. This reflects the strong CO--C interaction energy of \(1600~\mathrm{K}\) used in the simulations (see Appendix A).

At monolayer completion, the molecular clusters observed by STM as bright protrusions are not reproduced by the KMC simulation; only transient second-layer adsorbates are observed (Fig.~\ref{fig3}E). In the model, adsorption is favored at uncoordinated sites, so that the absence of such stabilizing sites on a perfect, defect-free monolayer kinetically hinders second-layer growth. However, STM observations show that even when bright protrusions form, they do not lead to sustained second-layer growth. Regime~II therefore cannot be explained solely by the scarcity of nucleation defects. Instead, as discussed below, it points to an orientational mismatch between the first and second CO layers.

\section{Discussion}

Classical descriptions of adsorption on interstellar grains originate in early kinetic models (e.g. in \cite{Watson1972, Williams1968}). In these treatments, the capture probability \(P_{\mathrm{cap}}\) at impact is assumed to approach unity for heavy species other than H or D \citep{Al-Halabi2003, Al-Halabi2004}, because their kinetic energy is rapidly dissipated into the grain within picoseconds after impact. Consistently, molecular beam experiments show that CO dissipates nearly all of its incident kinetic energy, largely independently of gas temperature \citep{Kohrt1973}. Thus, for heavy species like CO, there is no residual translational energy to sustain any post-impact motion on the surface. In addition, because thermal desorption from a stabilized adsorbed state is negligible at dust temperatures (\(\sim10~\mathrm{K}\)), heavy adsorbed species are assumed to be essentially immobile. Thus, astrochemical models treat their capture as effectively irreversible, and their sticking probability is then taken to be unity. However, this hit-and-stick assumption fails to explain low sticking and high mobility on flat surfaces at \(10~\mathrm{K}\), as well as delayed ice growth on grain analogues. An extended framework is therefore required, in which post-impact dynamics play a central role in determining whether adsorption becomes stable.

Surface science has long recognized that effective energy dissipation at capture does not by itself guarantee stabilization. After capture, molecules may still undergo thermally coupled surface exploration in a mobile precursor state and either locate a stabilizing site, such as islands or defects, or return to the gas phase \citep{Ehrlich1956, King1975, King1974, Kisliuk1957, Venables1984}. Although originally developed for high-temperature adsorption on metals, this mobile precursor framework reflects a more general physical idea: retention is not determined by immediate stabilization after capture, but by competition between stabilization and escape during an exploratory phase \citep{Kramers1940}. In low-temperature physisorption regimes, weak adsorbate--surface interactions can allow substantial post-impact mobility, so the same physical picture can be extended to this case: a mobile adsorbate may desorb, or relax without a substantial activation barrier into a stabilized physisorbed state. This defines a transient exploratory regime -- whose outcome is temperature dependent (Appendix A). This regime is distinct from the later bound-state regime reached once an adsorbate is trapped in a deep stabilizing site. The adsorbate--substrate binding energy of this final stabilized state is generally inferred from thermal desorption measurements \citep{Minissale2022}. In this bound-state regime, thermally activated motion during annealing can enable clustering \citep{Kouchi2020}, as also observed in our STM experiment (Fig.~\ref{fig:co_annealing}).

For CO on HOPG, the competition between stabilization and escape during the exploratory phase evolves continuously with exposure, governing how adsorption proceeds across the three regimes. At low exposures (Regime~I), stabilization is limited to step edges and other rare sites, while extended terraces remain largely devoid of stabilized monomers. Consistent with the statistical description given Appendix A, after capture, adsorbates perform many exploratory hops before stabilization, and even a small escape probability per hop leads to substantial desorption, resulting in the observed low sticking value of \(\sim0.3\). As islands expand, their edges provide additional stabilizing sites, and \(S\) rises toward \(\sim0.5\). In Regime~II, \(S\) decreases to \(\sim0.35\); at this stage, most captured molecules desorb and only a minority are retained on top of the completed monolayer, but a continuous second layer does not wet. As reported for CO on graphite \citep{Heidberg1990, Larher1987} and for various physisorbed films \citep{Larher1968, Larher1977, Terlain1983, Thomy1994}, such incomplete wetting is attributed to structural mismatch between the first and second layer, including positional and orientational effects. In the \(\sqrt{3}\) structure, CO lies flat in a geometry incompatible with the tilted, close-packed arrangement of the \(\alpha\)-CO(111) plane \citep{You1985}. Forming a second layer thus requires reorientation toward the (111) motif and angular matching for epitaxial growth. A similar Regime~II is observed for CO (and \ch{N2}), on water ice and olivine substrates (Figs.~\ref{fig:three_regime_substrates}--\ref{fig:co_n2_olivine}), suggesting that this bottleneck arises from first-layer self-organization rather than substrate-specific bonding \citep{Bienfait1985}.

At higher exposure (Regime~III), repeated adsorption attempts eventually allow enough molecules to access stabilizing second-layer configurations for three-dimensional growth to start. The sticking coefficient rises sharply toward \(\sim0.6\) as the system transitions to a Stranski--Krastanov growth mode \citep{Venables1984}, consistent with similar CO island growth reported on non-porous amorphous water ice \citep{Kouchi2021}. That \(S\) remains well below unity in this regime suggests that the Stranski--Krastanov growth pathway established at early stages continues to hinder multilayer growth. By promoting partially coalesced three-dimensional islands rather than a continuous wetting layer, this pathway delays the formation of an extended (111)-like molecular layer and the onset of efficient layer-by-layer sticking, consistent with previous observations that CO does not completely wet graphite during multilayer growth \citep{Larher1987}.

These three regimes have direct implications for modelling the freeze-out of CO and \(\mathrm{N}_2\) in cold, shielded environments from molecular clouds to protostellar regions \citep{Oberg2021}. Infrared observations indicate that a large fraction of interstellar CO ice is present in a nearly pure form \citep{Pontoppidan2003}, resulting from direct molecular condensation on grains. Our results show, however, that this growth cannot be described by a single constant sticking coefficient: for CO, adsorption passes through a reduced-retention regime near monolayer completion, before multilayer ice growth resumes with sticking still below unity over the first few layers. The similarly reduced sticking measured for \(\mathrm{N}_2\) on realistic grain analogues (Fig.~\ref{fig:co_n2_olivine}) \citep{Stadler2024} suggests that retention-limited freeze-out may also shape the build-up of \(\mathrm{N}_2\) ices on cosmic dust.\\

Our experimental results indicate that the sticking coefficient should not be treated as a fixed parameter, but as the outcome of a kinetic competition between desorption and retention occurring during post-impact exploration. To this end, we define \(P_{\mathrm{ret}}^0\) as the effective retention probability on a planar reference, with corresponding sticking coefficient \(S_0 = P_{\mathrm{cap}}P_{\mathrm{ret}}^0\). For heavy species, the capture probability at impact \(P_{\mathrm{cap}} \approx 1\), so \(S_0\) is directly constrained by \(P_{\mathrm{ret}}^0\). \(S_0\) is an empirical quantity that must be determined experimentally, since it reflects the retention landscape specific to the planar system of reference and to a given temperature. We then introduce \(\Delta\mu\) as an additional free-energy shift that modifies the balance between desorption and retention relative to that reference state. Such a shift may arise, for example, from additional adsorption sites (e.g. defects, ad-atoms), entropic constraints, external fields, curvature, etc. Under thermal coupling to the surface bath, the relative probability of configurations differing by \(\Delta\mu\) follows an effective Boltzmann factor \(\exp(-\Delta\mu/k_{\mathrm{B}}T)\). The retention probability then follows

\begin{equation}
P_{\mathrm{ret}} = P_{\mathrm{ret}}^0
\exp\left(-\frac{\Delta\mu}{k_{\mathrm{B}}T}\right).
\label{eq:pret_deltamu}
\end{equation}

Additional stabilization is favored for \(\Delta\mu < 0\), whereas escape is favored for \(\Delta\mu > 0\). The corresponding sticking coefficient is

\begin{equation}
S = S_0
\exp\left(-\frac{\Delta\mu}{k_{\mathrm{B}}T}\right).
\label{eq:sticking_deltamu}
\end{equation}

Negative \(\Delta\mu\) can strongly enhance \(S\), with the physical constraint \(S \leq 1\).\\

On soot, morphology provides such additional free-energy contribution. It can be described by the classical Kelvin relation,

\begin{equation}
\Delta\mu_{\mathrm{morph}}(r) =
\frac{2\gamma v_{\mathrm{m}}}{r},
\label{eq:deltamu_morph}
\end{equation}

where \(\gamma\) is an effective surface free energy associated with the curved interface, \(v_{\mathrm{m}}\) is the molecular volume, and \(r\) is the local radius of curvature. Concave regions (\(r < 0\)) favor stabilization, whereas convex regions (\(r > 0\)) promote escape. 

Substituting this free-energy shift into the general expression of \(S\) yields

\begin{equation}
S(r) = S_0
\exp\left(-\frac{2\gamma v_{\mathrm{m}}}{r k_{\mathrm{B}}T}\right).
\label{eq:sticking_curvature}
\end{equation}

Although negligible for micron-scale grains \citep{Papoular2005}, curvature effects become substantial at nanometer radii, lowering effective sticking and shifting the onset of adsorption to higher exposures relative to planar surfaces. Tolman corrections to \(\gamma\) are discussed in Appendix B.

Equation~\eqref{eq:sticking_curvature} quantifies how nanometer-scale convex curvature (\(r > 0\)) suppresses sticking. The corresponding suppression factors \(S/S_0\) for CO and \(\mathrm{N}_2\) at \(10~\mathrm{K}\) are summarized in Table~1. Both species are key molecules in cold astrophysical environments, and their similar volatilities make them a natural pair for illustrating how nanoscale curvature can bias retention on cosmic dust. The values listed in Table~1 reveal the strong sensitivity of retention to convexity: at interstellar temperatures, even modest positive curvature suppresses sticking by orders of magnitude relative to a flat surface.

\begin{table}
\caption{
Curvature-induced suppression factor \(S/S_0\) for convex surfaces (\(r>0\)).
The morphology-dependent suppression factor \(S/S_0\) is calculated from
\(S/S_0 = \exp[-2\gamma v_{\mathrm{m}}/(r k_{\mathrm{B}}T)]\)
for CO and \(\mathrm{N}_2\) at \(10~\mathrm{K}\).
For CO, we use \(\gamma = 12~\mathrm{mJ\,m^{-2}}\) and
\(v_{\mathrm{m}} = 5.1 \times 10^{-29}~\mathrm{m^3\,molecule^{-1}}\);
for \(\mathrm{N}_2\), we use \(\gamma = 10~\mathrm{mJ\,m^{-2}}\) and
\(v_{\mathrm{m}} = 5.8 \times 10^{-29}~\mathrm{m^3\,molecule^{-1}}\).
The surface tensions are based on cryogenic data \citep{Sprow1966},
with a \(10\)--\(20\%\) upward correction to approximate the solid phase,
and molecular volumes are taken from condensed-phase values.
A Tolman correction, \(\delta_{\mathrm{T}} = 0.2~\mathrm{nm}\), is applied for
\(r \leq 2~\mathrm{nm}\) (Appendix B).
The planar surface corresponds to \(S/S_0 = 1\).
}
\label{tab:curvature_suppression}
\centering
\begin{tabular}{lcc}
\hline\hline
Radius of curvature \(r\) & CO & \(\mathrm{N}_2\) \\
\hline
\(r = +1~\mathrm{nm}\)  & \(1.7 \times 10^{-3}\) & \(9.2 \times 10^{-3}\) \\
\(r = +2~\mathrm{nm}\)  & \(2.3 \times 10^{-2}\) & \(6.5 \times 10^{-2}\) \\
\(r = +5~\mathrm{nm}\)  & \(0.17\) & \(0.27\) \\
\(r = +10~\mathrm{nm}\) & \(0.42\) & \(0.52\) \\
\(r = +20~\mathrm{nm}\) & \(0.63\) & \(0.72\) \\
Flat, \(r=\infty\)      & \(1\) & \(1\) \\
\hline
\end{tabular}
\end{table}

Convex nanoscale curvature thus emerges as a major source of sticking suppression on rough grains, whereas concave regions (\(r < 0\)) generate a negative free-energy bias that can locally favor retention. As detailed in Appendix B, however, the low sticking measured on soot shows that thermodynamic stabilization by concave curvature is not by itself sufficient to ensure stable adsorption. In a purely static morphological picture, if concave and convex curvatures occupied comparable surface fractions (Fig.~\ref{fig:soot_tem}) and if retention were nearly complete throughout the concave fraction but negligible on convex asperities, the mean sticking would remain on the order of the concave surface fraction, i.e. \(\sim0.5\). The observed value, \(\sim0.03\)--\(0.1\), is much lower, showing that most adsorption trajectories still escape before reaching, or stabilizing within, the most favorable concave basins. Even in thermodynamically favorable concave pockets, sticking is therefore dominated by post-impact dynamics, which biases mobile adsorbates toward escape rather than stabilization.

The relative roles of nanoscale morphology and surface chemical composition can be assessed by comparing chemically distinct grain analogues. Our measurements of CO and \(\mathrm{N}_2\) adsorption on microparticulate olivine and soot reveal comparable sticking responses despite their different compositions (Fig.~\ref{fig:three_regime_substrates}) \citep{Stadler2024}. Earlier XPS studies likewise reported low sticking for \(\mathrm{H}_2\mathrm{O}\) and \(\mathrm{CO}_2\) on realistic grain analogues, whether bare or pre-covered with water ice \citep{Laffon2021}. This indicates that, in the low-temperature physisorption regime, nanoscale structure can become more important than surface chemical composition in determining sticking.

This model therefore provides a unified physical picture of low-temperature sticking. By emphasizing post-impact dynamics, it explains the high mobility and reduced sticking observed on a flat, weakly binding surface such as graphite, accounts for the persistence of bare nanoscale convex regions on rough grains, and shows that adsorption can remain dynamically limited even in strongly concave environments where retention is thermodynamically favored. In this regime, surface chemical composition becomes secondary to the dynamical and morphological control of retention.

\section{Conclusions}
Our results provide a refined picture of how ice forms on cosmic dust across molecular clouds, protostellar envelopes, and protoplanetary disks. Gas--surface interactions on realistic dust grains emerge as processes controlled not by impact capture alone, but by post-impact exploration dynamics shaped by surface defects and nanoscale geometry, more than surface chemistry. This physical picture extends the classical ``hit-and-stick'' view of gas--grain interactions and can, in principle, apply to atoms, radicals, molecules, and ions. It provides a route to translate nanoscale surface morphology into effective sticking probabilities for astrochemical models. It explains why adsorption can remain inefficient even at 10 K, why chemically distinct grain analogues display similarly low sticking, and why ice growth proceeds through delayed, morphology-dependent stabilization rather than simple layer-by-layer accretion. It shows that sticking probabilities on grains should not be set to unity in astrochemical models.

This improved physical description of adsorption on grains also has direct implications for molecular simulations of gas-grain interactions, including the quantum-based calculations on which astrochemical modelling increasingly relies. Adsorption energies commonly inferred from temperature-programmed desorption do not characterize the shallow and weakly corrugated exploratory landscape sampled immediately after impact, whose local topology is shaped by surface defects, curvature, and coverage. Rather, they describe the final stabilized states of clusters, islands, or trapped adsorbates, where molecule--molecule interactions contribute together with molecule--surface binding. Using such deep wells as the potential-energy surface for isolated post-impact adsorbates would impose irreversible trapping by construction, artificially suppressing the transient mobility that controls retention, the exploration of potentially reactive surface sites, and the role of nanoscale morphology in selecting where stabilization can occur. Progress in quantum-based modelling of physisorption and sticking will therefore require separating entrance-channel dynamics from final stabilized adsorption states, while explicitly accounting for surface defects, local geometry, and coverage-dependent stabilization.

\begin{acknowledgements}
The authors thank G. Arthaud for technical support. We thank InterCat and Liv Hornekær (Aarhus University) for providing access to the LT-STM facility. C. S. was supported by a mobility fellowship from the AMUTech Institut (Aix-Marseille Université).
\end{acknowledgements}

\section*{Data availability}

The online movies associated with the kinetic Monte Carlo simulations are available as supplementary online material. Movie 1 corresponds to the flat-surface simulation shown in Fig.~3C--F. Movie 2 corresponds to the reduced \(E_{\mathrm{CO-C}} = 900~\mathrm{K}\) simulation discussed in Appendix~A.

\bibliographystyle{aa}

\bibliography{sources}

\
\clearpage

\pagebreak

\clearpage

   \FloatBarrier
\begin{appendix}
\nolinenumbers % arXiv: aa.cls re-enables line numbering at appendix start
\twocolumn
%____________________________________________________________
\section{Exploration--escape controlled sticking}
\label{app:exploration_escape}

\subsection{Physical picture of exploration--escape controlled sticking}
\label{app:physical_picture_exploration_escape}

\begin{figure*}[!hb]
\centering
\includegraphics[width=18cm, height=8cm]{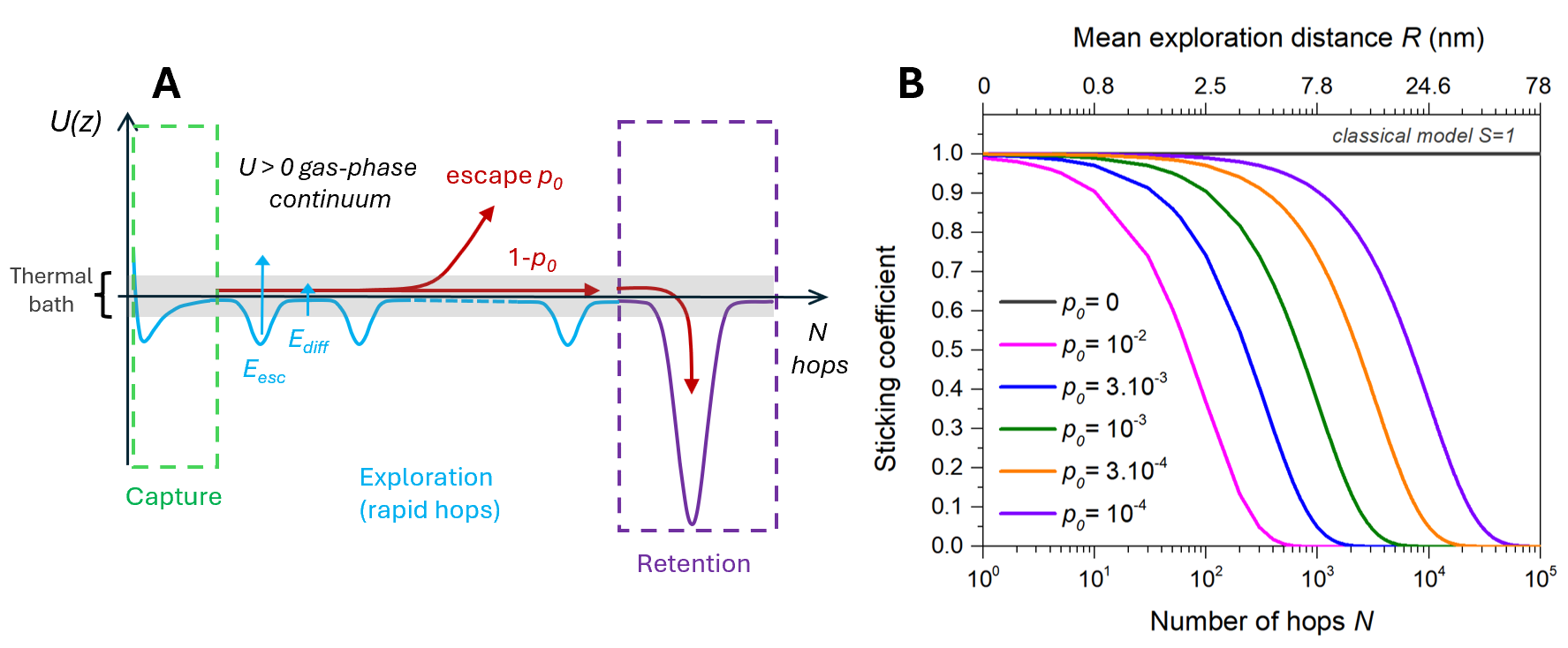}
\caption{
Exploration--escape controlled sticking of CO on HOPG.
(A) Schematic exploration--retention picture. After capture, the adsorbate explores the surface through rapid hops between shallow wells while remaining weakly bound and coupled to the substrate thermal bath. Lateral motion occurs over diffusion barriers, whereas thermal fluctuations may promote escape to the gas phase over an effective escape barrier, with an elementary escape probability per hop. Retention occurs only if a stabilizing configuration is reached before escape.
(B) Illustrative dependence of the sticking coefficient on the number of exploratory hops, highlighting the strong sensitivity of retention to small escape probabilities.
}
\label{fig:exploration_escape}
\end{figure*}

After initial capture, the adsorbate evolves within a mobile state and explores the shallow potential landscape of the terraces (Figure~\ref{fig:exploration_escape}). It diffuses through thermally activated hops between weakly bound configurations, progressively sampling the local environment. Stabilization occurs when the adsorbate reaches deeper binding configurations, such as existing islands or intrinsic surface features \citep{Venables1984}. Escape can occur at any stage of this exploration if thermal fluctuations enable crossing of the barrier toward the gas-phase continuum (\(U>0\)), thereby terminating the adsorption event.

\FloatBarrier

Within a transition-state framework, both escape and lateral motion correspond to thermally activated barrier crossings on a potential energy surface \citep{Ehrlich1956}. The escape rate from the shallow exploratory minima can be written in the standard transition-state form
\begin{equation}
k_{\mathrm{esc}} \approx \nu_{\mathrm{esc}}
\exp\left(-\frac{E_{\mathrm{esc}}}{k_{\mathrm{B}}T}\right),
\label{eq:kesc}
\end{equation}
where \(E_{\mathrm{esc}}\) is the effective depth of the exploratory minima relative to the desorption continuum. Lateral motion between adjacent shallow wells occurs with rate
\begin{equation}
k_{\mathrm{hop}} \approx \nu_{\mathrm{hop}}
\exp\left(-\frac{E_{\mathrm{diff}}}{k_{\mathrm{B}}T}\right),
\label{eq:khop}
\end{equation}
where \(E_{\mathrm{diff}}\) is the saddle-point barrier between neighboring exploratory minima.

STM observations on graphite indicate terrace-scale mobility prior to stabilization, implying that adsorbates explore lateral distances of order \(R \sim 10\)--\(100~\mathrm{nm}\) before becoming trapped. In a two-dimensional random walk with step length \(a\), the mean squared displacement satisfies \(\langle R^2\rangle = Na^2\), so that probing a lateral scale \(R\) requires \(N \sim (R/a)^2\) hops. Using \(a = 2.46~\mathrm{\AA}\) for graphite, exploration over \(10\)--\(100~\mathrm{nm}\) involves \(N \sim 1.6 \times 10^3\)--\(1.6 \times 10^5\) hops.

As an order-of-magnitude estimate for the transient post-impact regime, this exploration should occur on sub-second timescales; otherwise, one would expect a population of slowly mobile adsorbates on graphite terraces rather than rapid trapping at stabilizing sites. With \(\nu_{\mathrm{hop}} \sim 10^{12}\)--\(10^{13}~\mathrm{s^{-1}}\), sub-second exploration over these length scales at \(10~\mathrm{K}\) requires \(k_{\mathrm{hop}} \gtrsim 10^5\)--\(10^7~\mathrm{s^{-1}}\), corresponding to diffusion barriers \(E_{\mathrm{diff}} \sim 140\)--\(170~\mathrm{K}\), that is, \(\sim12\)--\(15~\mathrm{meV}\), consistent with motion within shallow exploratory wells. Because the diffusion barrier represents only a fraction of the escape barrier, adopting the conservative ratio \(E_{\mathrm{diff}}/E_{\mathrm{esc}} \approx 0.3\) implies \(E_{\mathrm{esc}} \sim 466\)--\(566~\mathrm{K}\).

These exploratory barriers are comparable in order of magnitude to diffusion barriers of \(\sim200~\mathrm{K}\) inferred for CO on non-porous amorphous water ice from TEM island-growth experiments for simple molecules on ice \citep{Furuya2022, Kouchi2020, Kouchi2021}. Although those experiments probe annealing regimes rather than the immediate post-impact regime, the measured diffusion lengths reflect monomer mobility before incorporation into molecular clusters, providing a useful order-of-magnitude check for the diffusion barrier in the transient mobile state. They further support the view that CO mobility is governed by weak substrate interactions of comparable magnitude on HOPG and water ice, consistent with the similarly reduced sticking coefficients and similar growth regimes measured on HOPG and non-porous crystalline \(\mathrm{H}_2\mathrm{O}\) ice (Fig.~\ref{fig:three_regime_substrates}), including the structural bottleneck that follows completion of the first CO layer .

This exploratory dynamics can be formulated statistically. At each step, the adsorbate faces two competing activated pathways: lateral hopping to a neighboring shallow well, with rate \(k_{\mathrm{hop}}\), or escape toward the gas phase, with rate \(k_{\mathrm{esc}}\). In this two-channel picture, the elementary probability of escape per step \(p_0\) is therefore
\begin{equation}
p_0 \simeq \frac{k_{\mathrm{esc}}}
{k_{\mathrm{esc}} + k_{\mathrm{hop}}},
\label{eq:p0}
\end{equation}
while the complementary probability \(1-p_0\) corresponds to continued exploration. If the adsorbate performs \(N\) hops, each associated with an elementary escape probability \(p_0\), the statistical retention probability \(P_{\mathrm{ret}}^0\) introduced in Eq.~\eqref{eq:pret_deltamu} can be expressed as
\begin{equation}
P_{\mathrm{ret}}^0 \simeq (1-p_0)^N,
\label{eq:pret0}
\end{equation}
which highlights the cumulative effect of escape during exploration (Fig.~\ref{fig:exploration_escape}). Here, the exploratory thermodynamics is encoded both in \(N\), which reflects the trajectory length before stabilization, and in \(p_0\), which depends on the local balance between hopping and escape. Even for \(p_0 \ll 1\), large \(N\) leads to a substantial escape probability and thus to a decrease of the retention probability \(P_{\mathrm{ret}}^0\). For \(N \sim 10^3\), \(p_0 \sim 10^{-3}\) reduces sticking by about a factor of two, consistent with \(S \approx 0.5\) for CO on graphite at \(10~\mathrm{K}\).

\FloatBarrier

\subsection{Temperature dependence of the exploratory phase}
\label{app:temperature_dependence_exploration}

\begin{figure*}[!b]
\centering
\includegraphics[width=17cm, height=8cm]{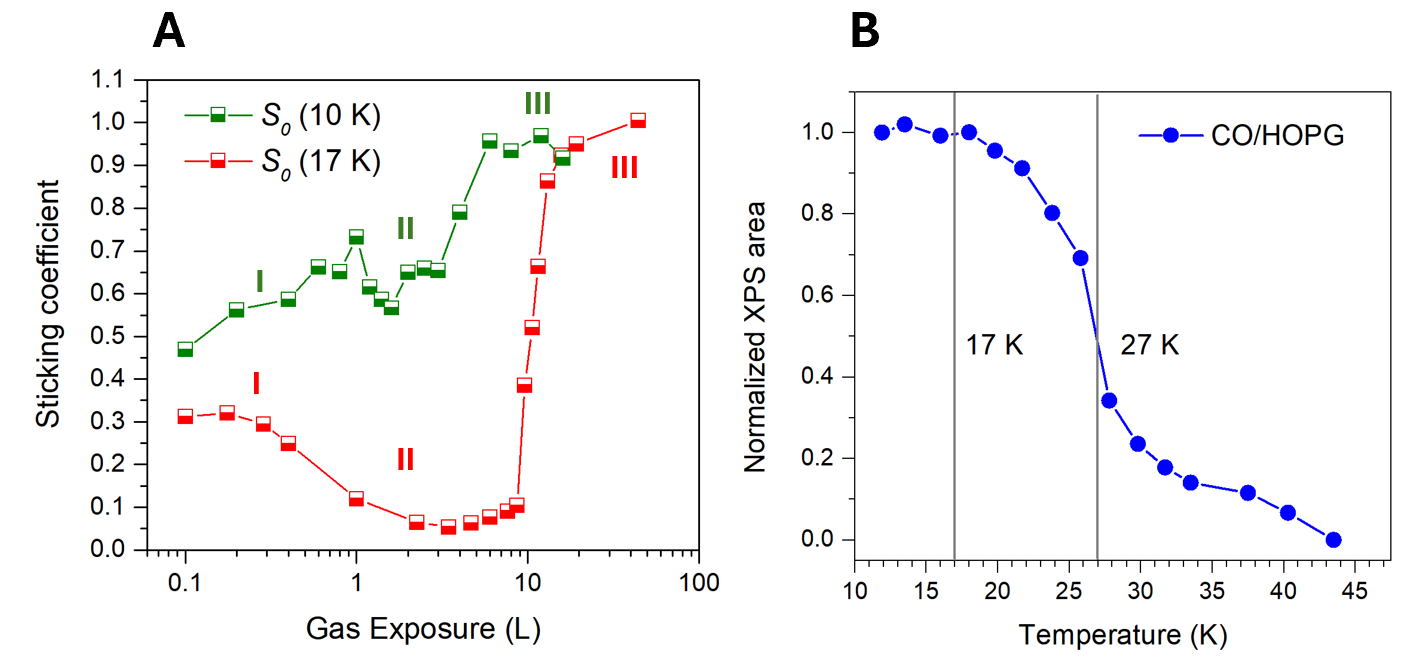}
\caption{
Temperature dependence of sticking of CO on HOPG.
(A) Experimental sticking coefficient of CO on HOPG measured by XPS at \(10~\mathrm{K}\) and \(17~\mathrm{K}\) as a function of gas exposure, showing strongly suppressed first-layer sticking at \(17~\mathrm{K}\), followed by an extended intermediate Regime~II and multilayer growth at higher exposure.
(B) Thermal desorption of CO from HOPG measured by XPS during linear heating, indicating an onset of desorption near \(27~\mathrm{K}\) and confirming that measurements at \(17~\mathrm{K}\) are well below the desorption regime.
}
\label{fig:temperature_dependence}
\end{figure*}

The temperature dependence shown in Figure~\ref{fig:temperature_dependence} provides strong evidence for the nanoscale mechanism governing the exploratory phase. At \(17~\mathrm{K}\), first-layer sticking of CO on HOPG is strongly reduced compared to \(10~\mathrm{K}\), even though both temperatures remain far below the desorption threshold of \(\sim27~\mathrm{K}\). This reduction can be understood as an increased probability of escape during the exploratory phase. Since \(p_0 \simeq k_{\mathrm{esc}}/(k_{\mathrm{esc}}+k_{\mathrm{hop}})\), the escape rate increases with temperature, so that a transiently captured molecule is more likely to desorb before reaching a stabilizing site.

\FloatBarrier

At \(17~\mathrm{K}\), Regime~II extends over a wider exposure range. This first reflects the lower sticking coefficient: more exposure is required to accumulate enough adsorbed CO to approach second-layer nucleation. In addition, the onset of multilayer growth is delayed by a configurational bottleneck. Stabilization of correctly oriented second-layer molecules becomes less favorable as temperature increases, owing to the larger entropic penalty in
\(\Delta G_{\mathrm{ads}} = \Delta H_{\mathrm{ads}} - T\Delta S_{\mathrm{ads}}\).
Because adsorption reduces the translational and rotational freedom of the molecule, \(\Delta S_{\mathrm{ads}} < 0\), and the opposing term \(-T\Delta S_{\mathrm{ads}}\) grows with temperature. As exposure increases and multilayer growth begins (Regime~III), stabilization becomes more probable, reducing the number of exploratory hops required before retention. The cumulative escape probability therefore decreases, and sticking abruptly recovers to \(\approx 0.6\), similar to the value at \(10~\mathrm{K}\), as expected below the onset of thermal desorption from stabilized adsorbed states.

Once CO is retained in deep wells, the system enters a distinct regime governed by thermally activated diffusion and desorption kinetics \citep{Kouchi2020}, as illustrated by clustering upon annealing (Fig.~\ref{fig:co_annealing}). These deep wells are characterized by temperature-programmed desorption measurements \citep{Minissale2022}, semi-empirical physisorption models \citep{Vidali1991}, or quantum-based calculations \citep{Rutigliano2021}. For CO on graphite, conventional adsorption energies are typically estimated to be around \(1600~\mathrm{K}\). If the exploratory landscape were controlled by this value, thermal diffusion at \(10~\mathrm{K}\) would be negligible, in contradiction with the terrace-scale mobility observed experimentally by STM, implying much shallower exploratory wells of only a few hundred kelvins. Kinetic Monte Carlo simulations support this picture: a shallower interaction energy, \(E_{\mathrm{CO-C}} = 900~\mathrm{K}\), restores terrace mobility, as shown in Movie~2, whereas a higher value, \(E_{\mathrm{CO-C}} = 1600~\mathrm{K}\), strongly suppresses diffusion.

A similar decoupling is observed for \(\mathrm{H}_2\mathrm{O}\) on HOPG. Although temperature-programmed desorption yields effective adsorption energies of \(\sim5700~\mathrm{K}\) \citep{Minissale2022}, values dominated by cooperative hydrogen-bond stabilization within ice aggregates, semi-empirical estimates for an isolated \(\mathrm{H}_2\mathrm{O}\) monomer on graphite are substantially lower, around \(\sim1900~\mathrm{K}\) \citep{Vidali1991}. Yet STM experiments show that water molecules remain highly mobile on graphite terraces at temperatures far below the desorption temperature of \(\sim155~\mathrm{K}\), diffusing over long distances before stabilizing at steps or defects \citep{Doktor2025}. Consistently, the sticking coefficient remains well below unity, with \(S \approx 0.4\) at \(20~\mathrm{K}\) \citep{Laffon2021}.

%____________________________________________________________
\section{Nanoscale curvature and sticking on soot}
\label{app:soot_curvature}

TEM images show that soot nanoparticle surfaces consist of alternating nanometric protrusions and depressions (Fig.~\ref{fig:soot_tem}). Each depression can be approximated as a circular arc of width \(w\) and depth \(d\), defining a local radius of curvature
\begin{equation}
r = \frac{w^2}{8d} + \frac{d}{2}.
\label{eq:soot_radius}
\end{equation}
For typical values measured on soot, \(w \approx 1.8~\mathrm{nm}\) and \(d \approx 0.5~\mathrm{nm}\), this yields \(|r| \approx 1.1~\mathrm{nm}\). At such radii, curvature modifies the effective surface free energy \(\gamma(r)\). This dependence is described by the Tolman correction \citep{Tolman1949}:
\begin{equation}
\gamma(r) = \frac{\gamma_0}{1 + 2\delta_{\mathrm{T}}/r},
\label{eq:tolman}
\end{equation}
where \(\delta_{\mathrm{T}}\) is the Tolman length. Kelvin-type thermodynamic relations have been experimentally validated down to radii of \(\sim0.5~\mathrm{nm}\) in capillary condensation, provided a correction of order \(\delta_{\mathrm{T}} \approx 0.2~\mathrm{nm}\) is included \citep{Kim2018}. Atomistic models similarly show deviations exceeding \(5\%\) below \(\sim5~\mathrm{nm}\) and recover Tolman-like behavior at leading order \citep{Wang2021}, supporting the validity of curvature-induced free-energy shifts at nanometer scales.

A planar surface free energy \(\gamma_0 \approx 12~\mathrm{mJ\,m^{-2}}\) is appropriate for solid CO, based on liquid values near \(70\)--\(80~\mathrm{K}\) \citep{Sprow1966}, applying a \(10\)--\(20\%\) increase for the solid phase. Using \(\delta_{\mathrm{T}} = 0.2~\mathrm{nm}\) and \(|r| \approx 1.1~\mathrm{nm}\) yields \(\gamma(r) \approx 8.8~\mathrm{mJ\,m^{-2}}\), corresponding to a substantial reduction relative to the planar value. In Table~\ref{tab:curvature_suppression}, a Tolman correction of \(\delta_{\mathrm{T}} = 0.2~\mathrm{nm}\) is applied for \(r \leq 2~\mathrm{nm}\).

\begin{figure}[!h]
\centering
\includegraphics[width=\columnwidth,height=6cm,keepaspectratio]{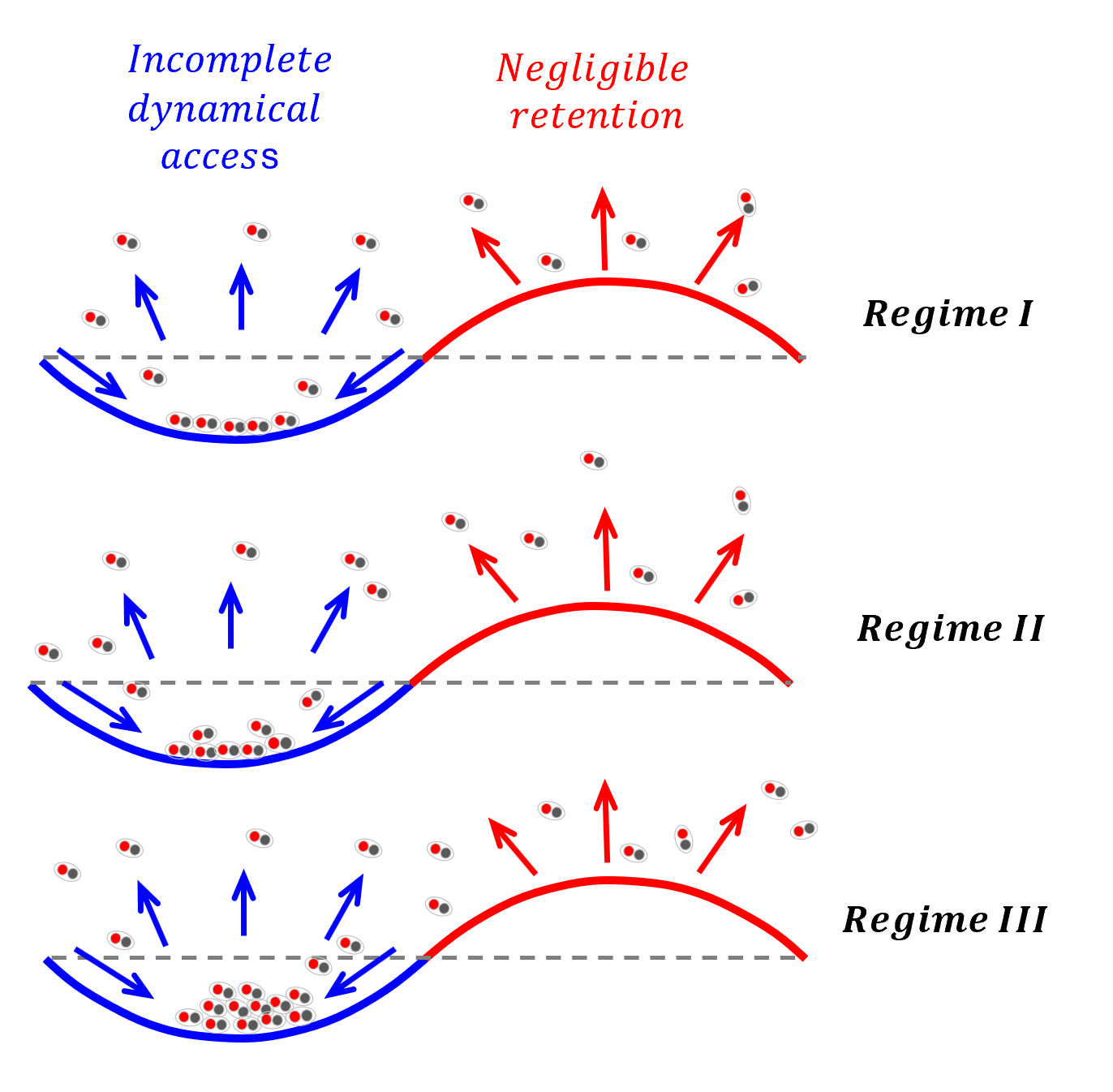}
\caption{
Schematic view of local retention on soot across the three adsorption regimes.
Concave pockets act as stabilizing regions for CO retention and nucleation - although dynamical access to stabilized sites is not complete -, whereas convex asperities contribute negligibly to retention. The three-regime evolution observed on HOPG is assumed to occur locally within these concave pockets: Regime~I corresponds to local retention in concave regions, Regime~II to reduced retention associated with the second-layer bottleneck, and Regime~III to enhanced retention as multilayer growth develops.
}
\label{fig:soot_retention}
\end{figure}

\begin{figure}[!htbp]
\centering
\includegraphics[width=\columnwidth,height=8cm,keepaspectratio]{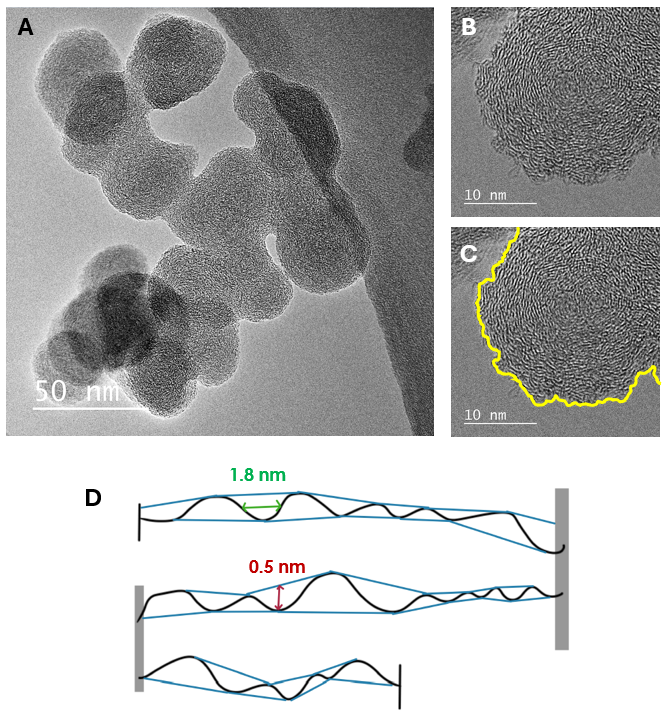}
\caption{
TEM characterization of soot aggregates showing hierarchical roughness at the nanometer scale.
(A) Representative soot aggregate composed of fused \(\sim25~\mathrm{nm}\) carbon nanoparticles.
(B) High-resolution image of an individual nanoparticle.
(C) Manual segmentation of the surface contour used to extract roughness profiles.
(D) Typical pore-depth \((0.5~\mathrm{nm})\) and pore-width \((1.8~\mathrm{nm})\) distributions, illustrating the nanometer-scale concave pockets and convex asperities that define the local morphology.
}
\label{fig:soot_tem}
\end{figure}

On soot, concave pockets and convex asperities occupy broadly comparable surface areas (Figure~\ref{fig:soot_retention}). Convex regions are expected to contribute negligibly to retention, so the measured sticking mainly arises from molecules that reach and stabilize within concave regions. A purely static morphological picture would then predict a mean sticking on the order of \(0.5\) if concave and convex regions occupied similar surface fractions and if local retention were nearly complete throughout the concave fraction but negligible on convex regions. The measured soot sticking, \(\sim0.03\)--\(0.1\), is far smaller, showing that this static picture strongly overestimates the effective stabilizing fraction of the surface.

This discrepancy indicates that most adsorption trajectories do not lead to stabilization, even though concave regions are locally favorable in the Kelvin description. In a realistic three-dimensional morphology, molecules transiently captured on rims, sidewalls, or weakly connected concave domains may continue to explore the surface and escape before becoming stabilized within a concave basin. The relevant quantity is therefore not simply the total concave area, but the fraction of post-impact trajectories that dynamically access and remain within stabilizing concave regions before escape. In addition, stabilization at low coverage may require local molecular organization. On graphite, retention in Regime~I is associated with formation of the commensurate \(\sqrt{3}\) phase. On a rough concave surface, local curvature and structural disorder may impose geometric or mechanical frustration that hinders and delays self-organization of this phase, even where the local free-energy bias favors retention, which may lower the adsorption probability in concave pockets.

The global sticking can be written schematically as
\begin{equation}
S_{\mathrm{total}} \approx \phi_{\mathrm{conc}}\, f\, S_{\mathrm{local}},
\label{eq:stotal_soot}
\end{equation}
where \(\phi_{\mathrm{conc}} \approx 0.5\) is the nominal concave surface fraction, \(S_{\mathrm{local}}\) is the local sticking probability within stabilizing concave regions, and \(f\) is an effective residual factor accounting for the reduction of the nominal concave contribution. This factor reflects incomplete dynamical access to favorable basins during post-impact exploration, as well as possible frustration of the ordered first-layer structures required for efficient stabilization.

Because soot and HOPG display the same three-regime evolution, we infer that the same underlying stabilization bottlenecks remain operative on both substrates, with the regime dependence carried by \(f\). Taking \(S_{\mathrm{local}} \sim 1\) within favorable concave regions and \(S_{\mathrm{local}} \sim 0\) on convex regions gives \(S_{\mathrm{total}} \approx 0.5f\). Matching the measured sticking coefficients on soot then requires \(f \sim 0.06\)--\(0.14\) in Regime~I, \(f \sim 0.06\)--\(0.10\) in Regime~II, and \(f \sim 0.12\)--\(0.18\) in Regime~III. Equivalently, only about \(6\)--\(18\%\) of the trajectories associated with the nominal concave fraction lead to effective stabilization within concave pockets.

\FloatBarrier

%____________________________________________________________

\clearpage

\FloatBarrier
\section{Additional experimental figures}
\label{app:additional_figures}

\noindent This appendix gathers additional experimental figures supporting the
structural characterizations and the adsorption behavior
of CO and \(\mathrm{N}_2\) on graphite, soot, olivine, and water ice.

\suppressfloats[t]

\begin{figure}[!htbp]
\centering
\includegraphics[width=\columnwidth,height=7cm,keepaspectratio]{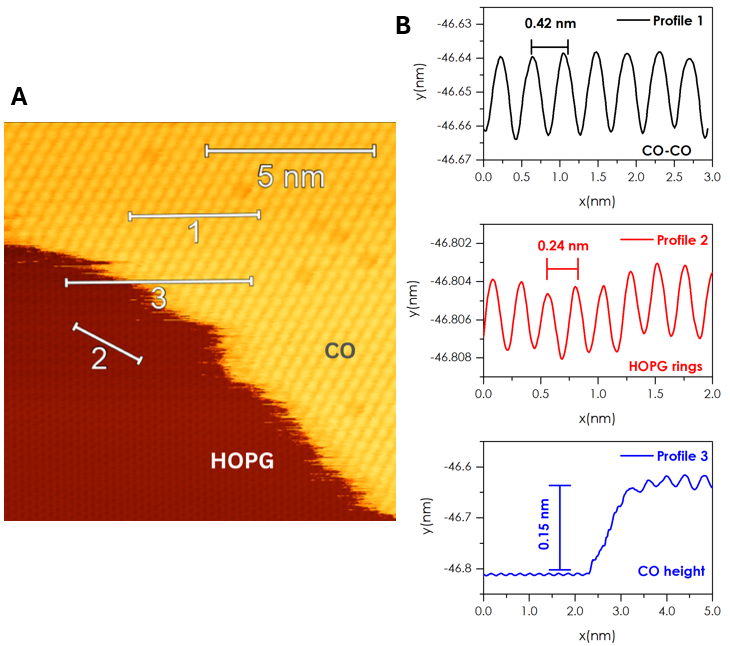}
\caption{
STM line scans of CO on HOPG.
(A) STM image of a CO island after a low CO dose (0.5 L).
White lines indicate the positions of the line profiles.
(B) Line profiles showing the commensurate \(\sqrt{3}\) phase, the HOPG lattice periodicity, and the apparent height of the CO monolayer.
}
\label{fig:stm_linescans}
\end{figure}

\begin{figure}[!htbp]
\centering
\includegraphics[width=\columnwidth,height=9cm,keepaspectratio]{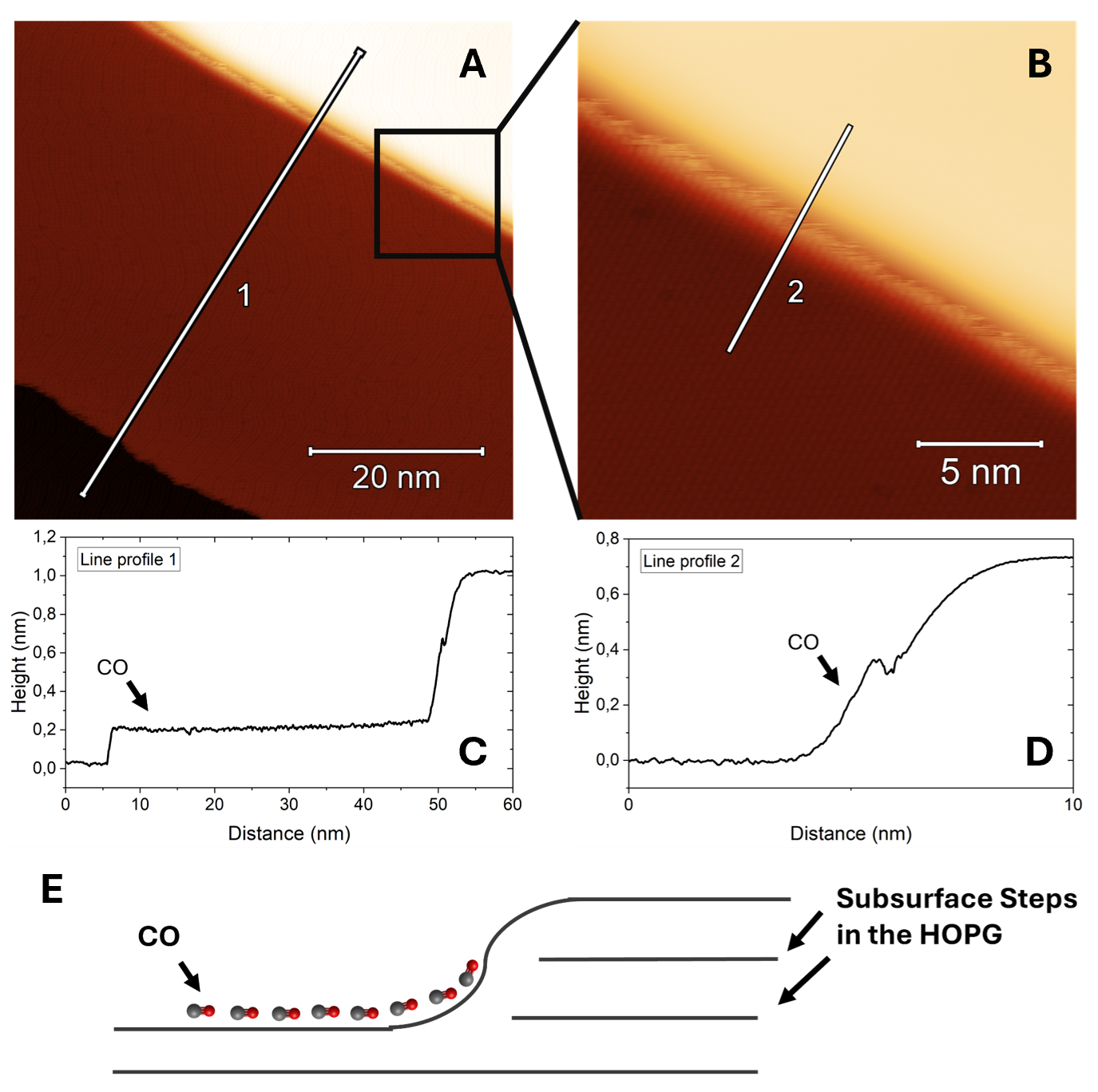}
\caption{
Morphology-driven preferential adsorption of CO on HOPG soft steps.
(A) STM image of a CO island on HOPG at low CO dose (0.5 L; scanning parameters I= 63 pA, V=-2.1 V), showing preferential accumulation along a smooth topographic modulation attributed to a subsurface step in the HOPG lattice.
(B) Zoomed view of the region highlighted in (A).
(C) Line profile across the island edge. The CO monolayer nucleates within the concave region of the soft step and expands laterally toward the flat terrace.
(D) Line profile across the soft step, showing gradual topographic curvature. The CO layer terminates where the curvature changes sign and becomes convex.
(E) Schematic illustration of preferential CO adsorption at a subsurface-induced soft step in HOPG.
}
\label{fig:stm_soft_steps}
\end{figure}

\begin{figure}[!htbp]
\centering
\includegraphics[width=\columnwidth,height=8cm,keepaspectratio]{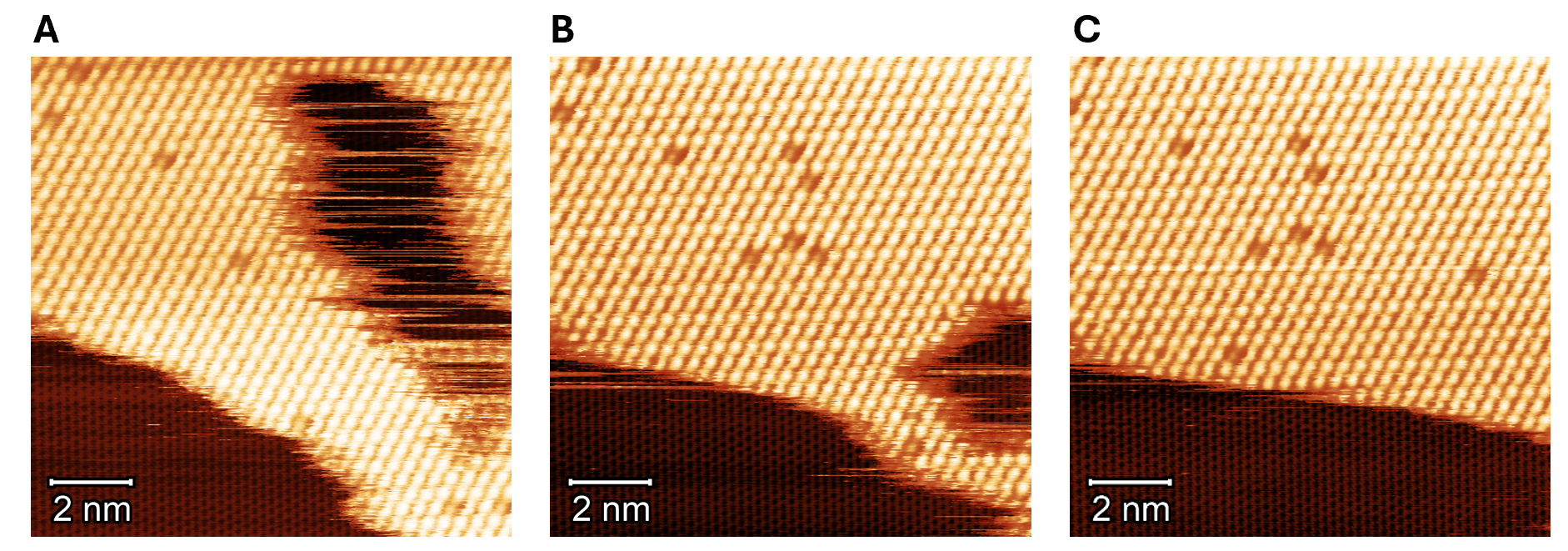}
\caption{
Tip-induced perturbation at \(4~\mathrm{K}\) and spontaneous reconstruction of a CO island on HOPG.
(A) STM image recorded after a deliberate tip-induced perturbation of the CO island, which displaced several molecules and created a local gap within the commensurate layer.
(B, C) Subsequent STM scans acquired under identical conditions, showing that CO molecules spontaneously rearrange to refill the gap, even at \(4~\mathrm{K}\).
The rapid reconstruction demonstrates weak CO--graphite interaction, consistent with the high lateral mobility of CO, and exploration governed by shallow binding on HOPG terraces.
Scanning parameters: \(I = 63~\mathrm{pA}\), \(V = -2.1~\mathrm{V}\).
}
\label{fig:stm_tip_reconstruction}
\end{figure}

\begin{figure}[!htbp]
\centering
\includegraphics[width=\columnwidth,height=8cm,keepaspectratio]{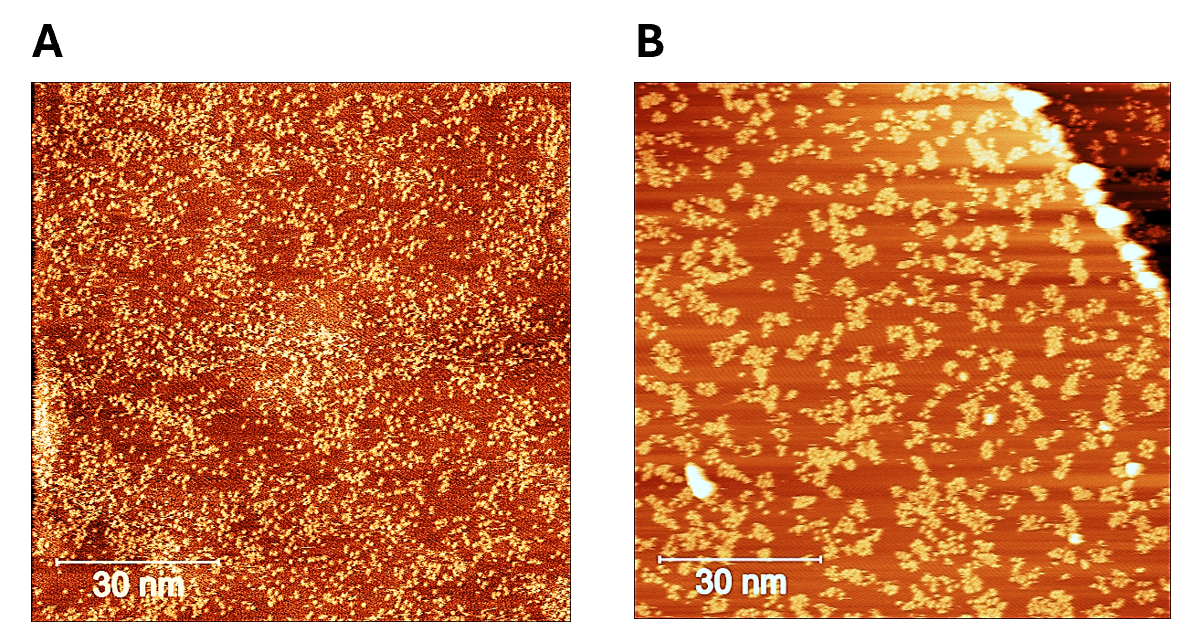}
\caption{
Annealing of a CO layer.
(A) STM image after dosing of \(5.4~\mathrm{L}\) of CO on HOPG at \(\sim 11~\mathrm{K}\), showing isolated second-layer nuclei across the first monolayer, reflecting the final stage of the transient exploratory regime, where CO is stabilized in deep potential wells.
(B) After annealing at \(25~\mathrm{K}\), thermally activated motion induces clustering, showing structural reorganization driven by thermally activated diffusion from these bound states.
Scanning parameters: \(I = 63~\mathrm{pA}\), \(V = -2.1~\mathrm{V}\).
}
\label{fig:co_annealing}
\end{figure}

\begin{figure}[!htbp]
\centering
\includegraphics[width=\columnwidth,height=8cm,keepaspectratio]{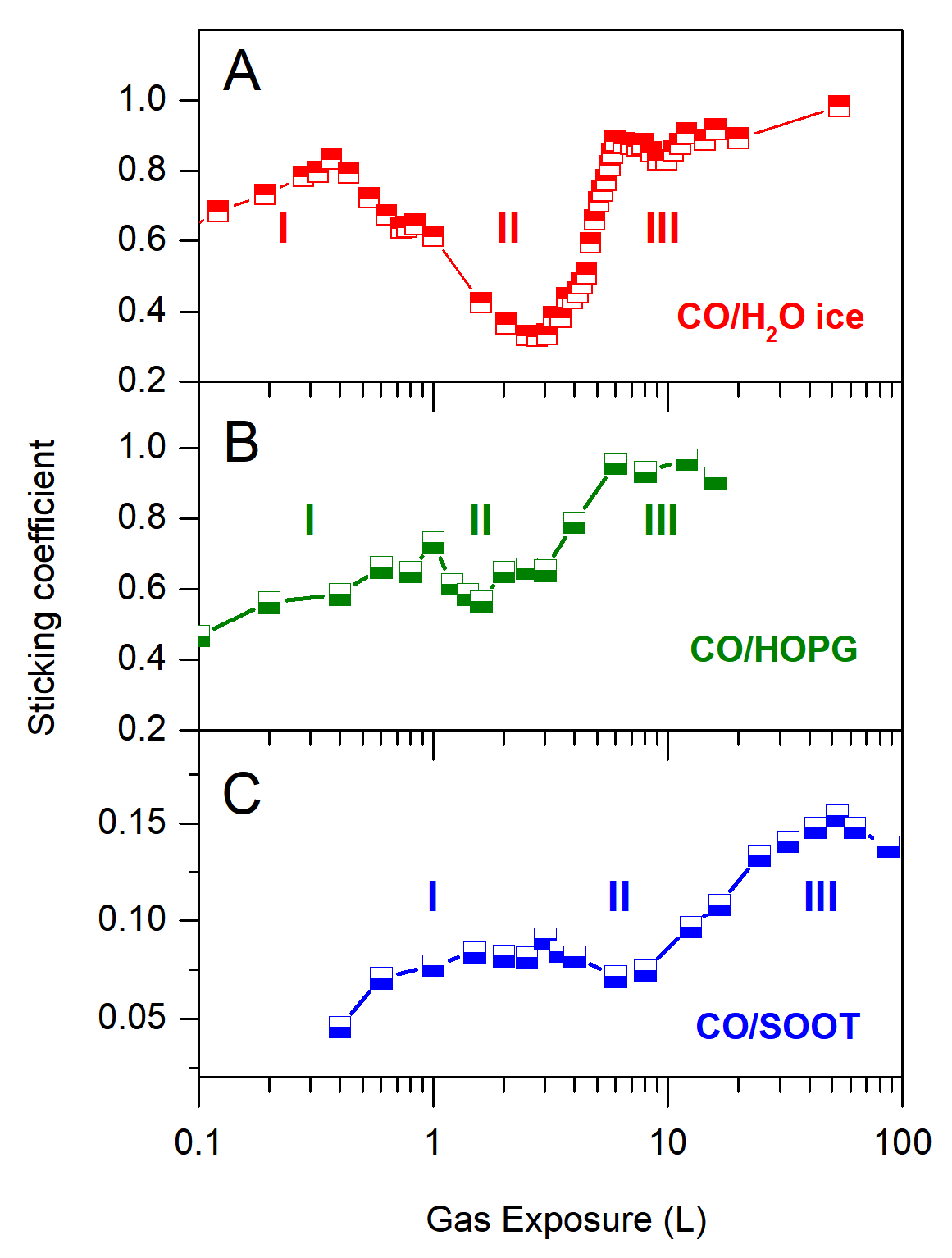}
\caption{
Three-regime sticking behavior of CO on different substrates.
Sticking coefficient of CO at \(10~\mathrm{K}\) as a function of gas exposure on
(A) non-porous crystalline \(\mathrm{H}_2\mathrm{O}\) ice,
(B) HOPG, and
(C) soot.
All substrates display the same adsorption regimes: Regime~I, reduced initial sticking; Regime~II, sticking reduced as first-layer completion is approached; and Regime~III, recovery toward multilayer growth.
Errors in \(S\) correspond to \(\pm 5\%\) for flat surfaces and \(\pm 10\%\) for particles.
}
\label{fig:three_regime_substrates}
\end{figure}

\begin{figure}[!htbp]
\centering
\includegraphics[width=\columnwidth,height=10cm,keepaspectratio]{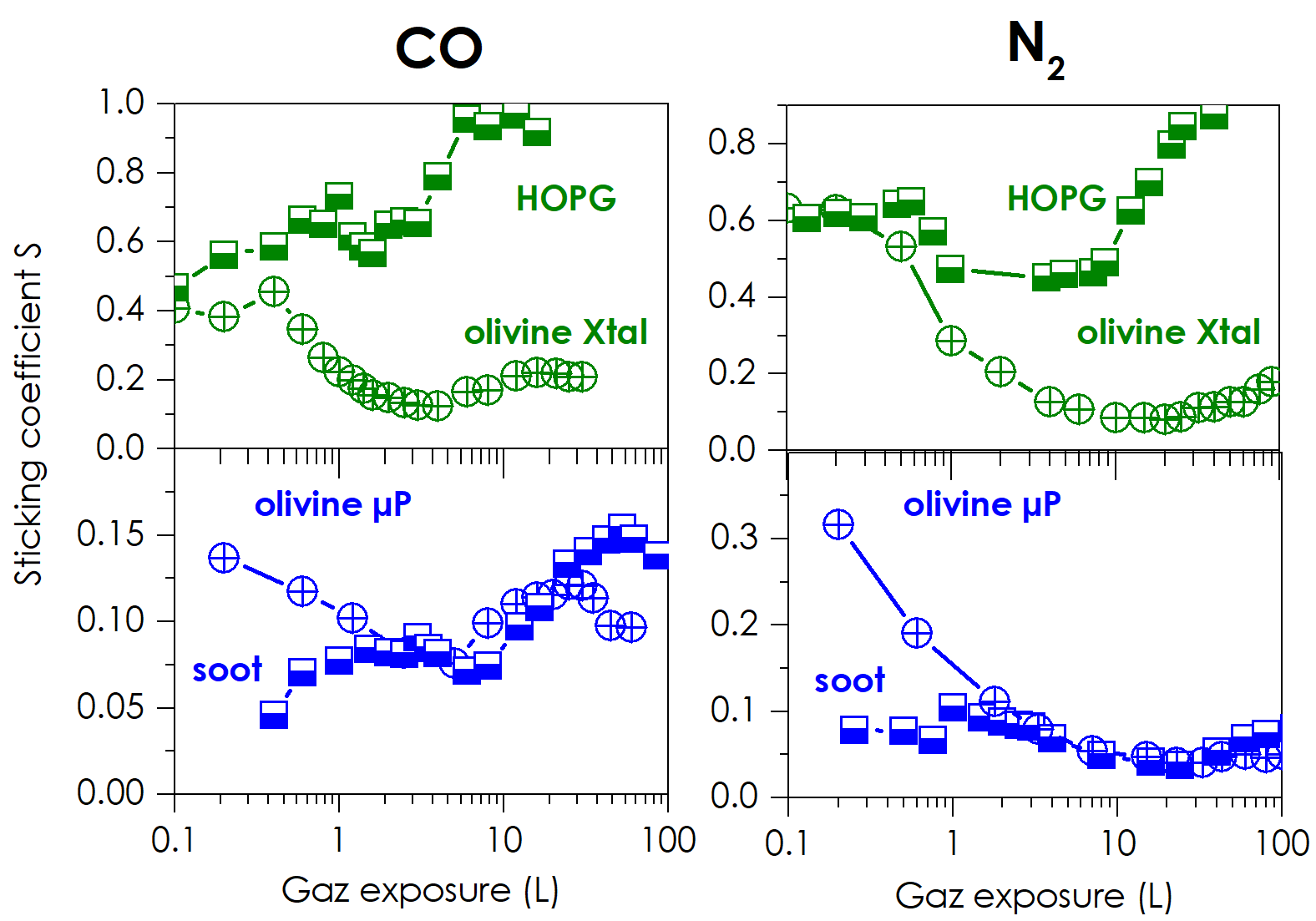}
\caption{
 Sticking coefficients of CO (left panel) and \(\mathrm{N}_2\) (right panel) extracted from XPS for HOPG, crystalline olivine, microparticulate olivine, and soot, showing comparable low and non-monotonic sticking behavior despite differences between CO and \(\mathrm{N}_2\).
These trends highlight the dominant role of nanoscale morphology in controlling sticking.
}
\label{fig:co_n2_olivine}
\end{figure}

\end{appendix}
\end{document}